\documentstyle{aipproc}

\input{epsf}

\def\Journal#1#2#3#4{{#1} {\bf #2}, #3 (#4)}

\def\NPB{{\em Nucl. Phys.} B}
\def\PLB{{\em Phys. Lett.}  B}
\def\PRL{\em Phys. Rev. Lett.}
\def\PRD{{\em Phys. Rev.} D}
\def\ZPC{{\em Z. Phys.} C}

\begin{document}

\title{Semiexclusive Processes: 
A different way to probe hadron structure\thanks{Invited talk at the
12th Nuclear Physics Summer School and Symposium: New Directions in
Quantum Chromodynamics, Kyongju, South Korea, 21--25 June 1999. }}

\author{Carl E. Carlson}

\address{Nuclear and Particle Theory Group, Physics Department,\\
College of William and Mary, Williamsburg, VA  23187-8795}

\maketitle

\vglue -5.3 cm  \hfill WM-99-116
\vglue 4.0cm

\begin{abstract}
Hard semiexclusive processes provide an opportunity to design effective
currents to probe specific parton distributions and to probe in leading
order parton distributions that fully inclusive reactions probe only
via higher order corrections. High transverse momentum pion 
photoproduction is an example of a such a process.  We
discuss the perturbative and soft processes that contribute, and show
how regions where perturbative processes dominate can give us the
parton structure information.  Polarized initial states are needed to
get information on polarization distributions.  Current polarization
asymmetry data is mostly in the soft region.  However, with somewhat
higher energy, determining the polarized gluon distribution using hard
pion photoproduction appears quite feasible.
\end{abstract}

\section*{Semi-Exclusive Processes as Probes of Hadron Structure}

This talk will discuss hard semiexclusive processes, namely
processes of the form $B+A\rightarrow C+X$, where the momentum
transfer $t = (p_B - p_C)^2$ is
large~\cite{many,peralta,cw93,acw97,acw98,bdhp99,b_epic,c_epic}. 
Both the cases where the hadron $C$ is part of a jet and where
it is kinematically isolated are interesting.  
Semiexclusive processes
provide the capability of designing ``effective
currents''~\cite{b_epic} that probe specific parton
distributions and for probing in leading order target
distributions that are not probed at all in leading order in inclusive
reactions.

Particle $B$ can be a hadron or a real or virtual photon.  We
will here limit ourselves to the latter.
The process we will discuss is 

\begin{equation}
\gamma + A \rightarrow M + X  ,
\end{equation}
where $A$ is the target and $M$ is a meson, for definiteness the
pion.  The process is perturbative because of the high transverse
momentum of the pion, not because of the high $Q^2$ of the photon.
Soft processes are from the present viewpoint an annoyance, but one we
need to discuss and we will estimate their size farther below. 

Our considerations also apply to electroproduction,

\begin{equation}
e+ A \rightarrow M + X
\end{equation}
when the final electron is not seen.  We use the
Weiz\"acker-Williams equivalent photon
approximation~\cite{bkt71} to relate the electron and photon
cross sections,

\begin{equation}
d\sigma(eA \rightarrow MX) = \int dE_\gamma \, N(E_\gamma) 
      d\sigma(\gamma A \rightarrow MX),
\end{equation}
where the number distribution of photons accompanying the electron is a
well known function.  

In the following section, we will describe the subprocesses that
contribute to hard pion production and show how the cross sections are
dependent upon the parton densities and distribution amplitudes
that we wish to probe, and in the subsequent section display some
results.  There will be a short summary at the end.

\section*{The Subprocesses}

\subsubsection*{At the Highest $k_T$}

\begin{figure}

\vskip 5mm

\centerline { \epsfxsize 3.5in \epsfbox{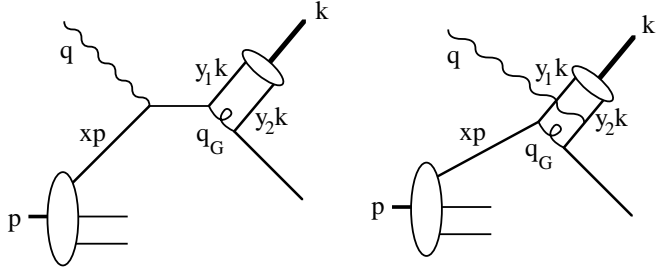}    }

\caption{Direct pion production.  The pion is produced in a  short
distance perturbatively calculated process, not by fragmentation of an
outgoing parton.  Thus this pion is kinematically isolated, and not
part of a jet.  Direct pion production gets important at very high
transverse momentum because the pion does not have to share the
available momentum with any other particles going in its direction. } 
\label{direct}

\end{figure}

At the highest possible transverse momenta, observed pions are directly
produced at short range via a perturbative QCD (pQCD) calculable
process~\cite{cw93,acw97,acw98,bdhp99}.  Two out of four lowest order
diagrams are shown Fig.~\ref{direct}.  The pion produced this way is
kinematically isolated rather than part of a jet, and may be seen
either by making  an isolated pion cut or by having some faith in the
calculation and going to a kinematic region where this process
dominates the others.  Although this process is higher twist, at the
highest transverse momenta its cross section falls less quickly than
that of the competition, and we will show plots indicating the
kinematics where it can be observed.


The subprocess cross  section for direct or short-distance pion
production is 

\begin{equation}
{d\hat\sigma \over dt} (\gamma q \rightarrow \pi^\pm q')
  = {128 \pi^2 \alpha \alpha_s^2 \over 27 (-t) \hat s^2}  I_\pi^2
   \left( {e_q \over \hat s} + {e_q' \over \hat u} \right)
 \left[  \hat s^2 + \hat u^2 + \lambda h (\hat s^2 - \hat u^2) \right] ,
\end{equation}
where $\hat s$, $\hat t = t$, and $\hat u$ are the subprocess
Mandlestam variables;  $\lambda$ and $h$ are the helicities of the
photon and target quark, respectively; and $I_\pi$ is the integral

\begin{equation}
I_\pi = \int {dy_1 \over y_1} \phi_\pi (y_1, \mu^2)  .
\end{equation}

\noindent In the last equation, $\phi_\pi$ is the distribution
amplitude of the pion, and describes the quark-antiquark part of the
pion as a parallel moving pair with momentum fractions $y_i$.  It is
normalized through the rate for $\pi^\pm \rightarrow \mu \nu$, and for
example,

\begin{equation}
\phi_\pi = {f_\pi \over 2 \sqrt{3}} 6 y_1 (1-y_1)
\end{equation}
for the distribution amplitude called ``asymptotic'' and for 
$f_\pi \approx 93$ MeV. Overall,

\begin{equation}
{d\sigma \over dx \, dt}(\gamma A \rightarrow \pi   X)
  = \sum_q  G_{q/A}(x,\mu^2) 
         {d\hat\sigma \over dt} (\gamma q \rightarrow \pi^\pm q')  ,
\end{equation}
where $G_{q/A}(x,\mu^2)$ is the number distribution for quarks of
flavor $q$ in target $A$ with momentum fraction $x$ at renormalization
scale $\mu$.

There are a number of interesting features about direct pion
production.

$\bullet$  For photoproduction, the struck quark's momentum
fraction  is fixed by
experimental observables.  This is like deep
inelastic scattering, where the experimenter can measure $x \equiv
Q^2/2m_N \nu$ and this
$x$ is also the momentum fraction of the struck quark (for high
$Q$ and $\nu$).  For the present case,  momenta are defined in
Fig.~\ref{direct} and the Mandlestam variables are

\begin{equation}
s = (p+q)^2; \qquad t = (q-k)^2; \quad {\rm and} \quad u = (p-k)^2.
\end{equation}
The Mandlestam variables are all observables, and the ratio

\begin{equation}
x = {-t \over s+u}
\end{equation}
is the momentum fraction of the struck quark.  We will let the reader
prove this.

$\bullet$ The gluon involved in direct pion production is well off
shell~\cite{cw93,acw97,acw98}.

$\bullet$  Without polarization, we can measure $I_\pi$, given
trust in the other parts of the calculation.  This $I_\pi$ is precisely
the same as the $I_\pi$ in both $\gamma^* \gamma \rightarrow \pi^0$
and 
$e\pi^\pm \rightarrow e \pi^\pm$ .

$\bullet$ We have polarization sensitivity.  For $\pi^+$ production at
high $x$,

\begin{equation}
A_{LL} \equiv {        \sigma_{R+} - \sigma_{L+} 
                                \over 
                       \sigma_{R+} + \sigma_{L+}          }
   =  {s^2 - u^2 \over s^2 + u^2  } \cdot  {\Delta u(x) \over u(x) }
\end{equation}
where $R$ and $L$ refer to the polarization of the photon, and $+$
refers to the target, say a proton, polarization.  Also, inside a $+$
helicity proton the quarks could have either helicity, and 

\begin{equation}
\Delta u(x) \equiv u_+(x) - u_-(x)  .
\end{equation}

The large $x$ behavior of both $d(x)/u(x)$ and 
$\Delta d(x)/\Delta u(x)$ are of current interest.  Most fits to the
data have the down quarks disappearing relative to the up quarks at
high $x$,  in contrast to pQCD which has definite non-zero predictions
for both of the ratios in the previous sentence.  Recent improved work
on extracting neutron data from deuteron targets, has tended to support
the pQCD predictions~\cite{wally}.  


\begin{figure}

\vskip 5mm

{\epsfxsize 2.8 in \epsfbox{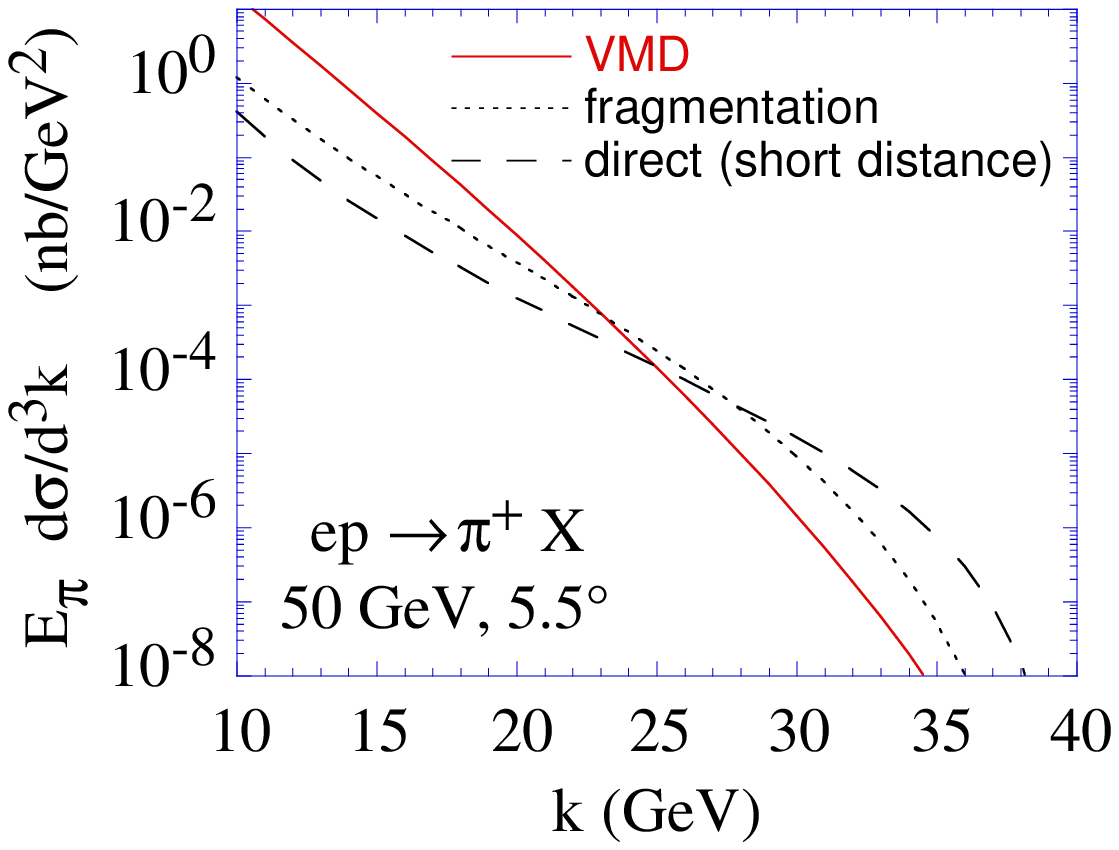}    }

{\vglue -2.22 in    \hglue 3in
\epsfxsize 2.8 in \epsfbox{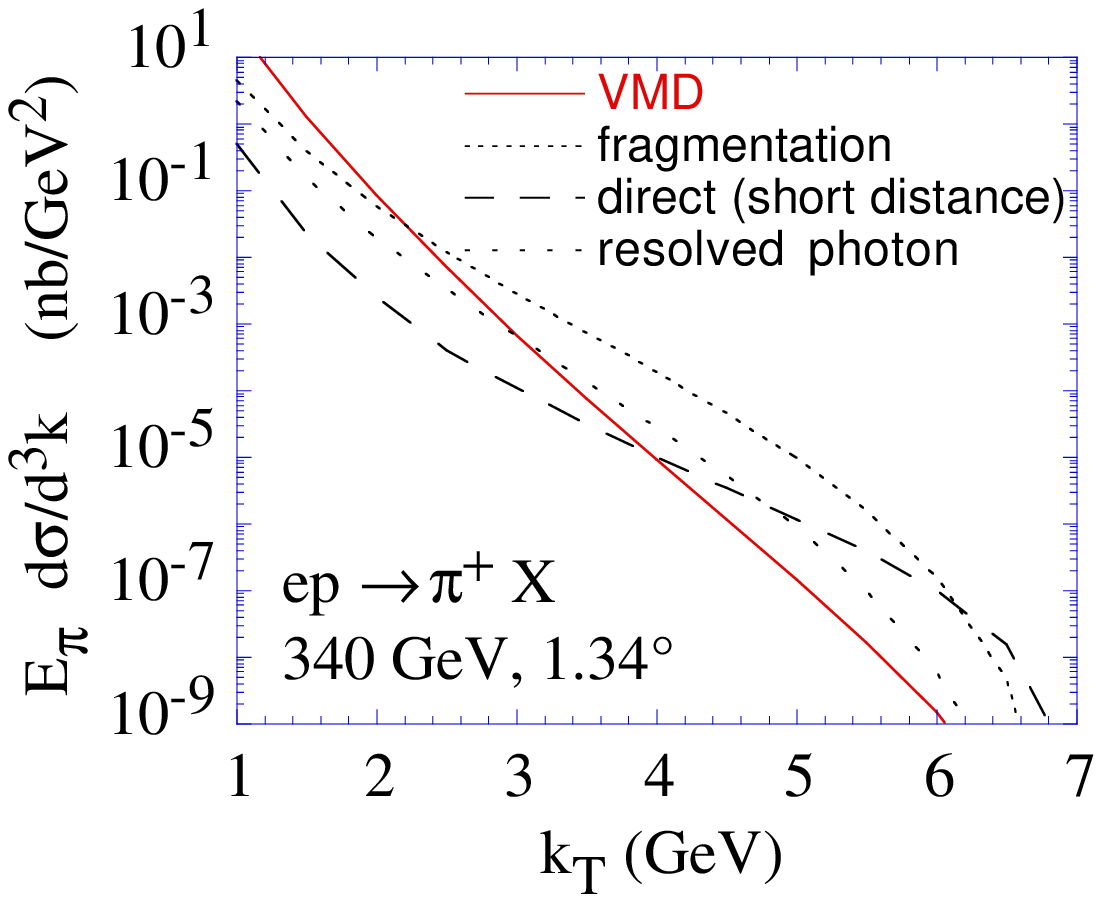}  }

\caption{Calculated contributions to the cross section for 
$ep \rightarrow \pi^+ X$, with electron beam energies and pions
emerging at angles in the target rest frame as indicated; $k$ is the
pion momentum.  The 50 GeV plot shows a significant region where direct
pion production dominates, once there is enough momentum that the soft
(``VMD'') processes are not large. The 340 GeV plot shows a long window
where the fragmentation process dominates.}   
\label{sigma}

\end{figure}


Experimentally, direct or short-range pion
production can be seen.  To show this, Fig.~\ref{sigma}(left) plots the
differential cross section for high transverse momentum $\pi^+$
electroproduction for a SLAC energy. Specifically, we have 50 GeV
incoming electrons, with the pion emerging at 5.5$^\circ$ in the lab.
It shows that above about 27 GeV total pion momentum or 2.6 GeV
transverse momentum,  direct (short distance, isolated) pion
production exceeds its competition.    Also shown in Fig.~\ref{sigma}
is a situation where there is a long region where the fragmentation
process---next up for discussion---dominates.  Incidentally, the 340
GeV energy for the electron beam on stationary protons was chosen to
match recent very preliminary discussions of an Electron Polarized Ion
Collider (EPIC) with 4 GeV electrons and 40 GeV protons,  and the
1.34$^\circ$ angle in the target rest frame matches 90$^\circ$ in the
lab for such a collider.  


\subsection*{Moderate $k_T$}

\begin{figure} 

\vskip 5mm

\centerline { \epsfxsize 3in \epsfbox{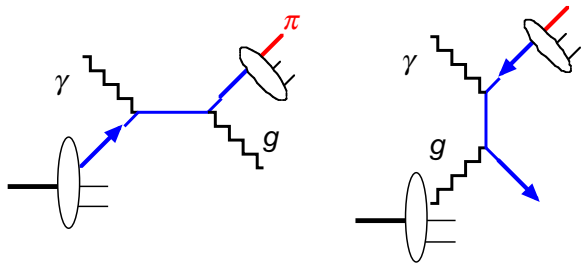}    }

\caption{The fragmentation process.  Pions are produced by
fragmentation of partons at long distances from the primary
interaction region.  The Compton process is on the left; the
pion could come from either quark or gluon fragmentation.  Quark-gluon
fusion is on the right.} 
\label{fragmentation}

\end{figure}

At moderate transverse momentum, the generally dominant process is
still a direct interaction in the sense that the photon interacts
directly with constituents of the target, but the pion is not produced
directly at short range but rather at long distances by fragmentation
of some parton~\cite{many,peralta,acw98}.  Many authors refer to this as
the direct process; others of us are in the habit of calling it the
fragmentation process.  The main subprocesses are called the Compton
process and photon-gluon fusion, and one example of each is shown in
Fig.~\ref{fragmentation}.


Photon gluon fusion often gives 30--50\% of the cross section for the
fragmentation process, and the polarization asymmetry is as large as
can be in magnitude,

\begin{equation}
\hat A_{LL}(\gamma g \rightarrow q \bar q) = -100\%.
\end{equation}

\noindent Typically for the Compton process, 
$\hat A_{LL}(\gamma q \rightarrow g q) \approx 1/2$.   We
shall show some $A_{LL}$ plots for the overall process after we
discuss the soft processes.

We should note that the NLO calculations for the fragmentation
process have been done also for the polarized case, though our plots
are based on LO.  For direct pion production,  NLO calculations are
not presently completed.

We should also remark that the photon may split into hadronic matter
before interacting with the target.  If splits into a quark anti-quark
pair that are close together, the splitting can be modeled
perturbatively or quasi-perturbatively, and we call it a ``resolved
photon process.''  A typical diagram is shown in the left hand part of
Fig.~\ref{hadronicphoton}. Resolved photon processes are crucial
at HERA energies, but not at energies under discussion
here, and we say no more about them.


\subsection*{Soft Processes}

\begin{figure} 

\vskip 5mm

\centerline { \epsfxsize 3.8in \epsfbox{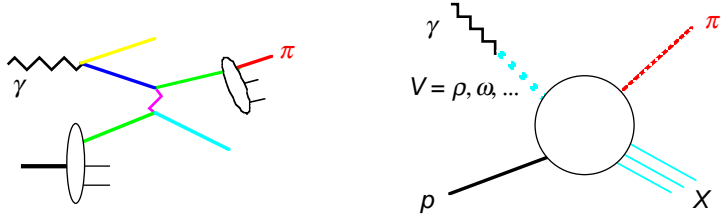}    }

\caption{Resolved photon process (left) and vector meson dominated
process (right).} 
\label{hadronicphoton}

\end{figure}

This is the totally non-perturbative part of the calculation, whose
size can be estimated by connecting it to hadronic cross sections.  The
photon may turn into hadronic matter, such as 
$\gamma \rightarrow q \bar q + \ldots$ with a wide spatial separation. 
It can be represented as photons turning into vector mesons. 
See Fig.~\ref{hadronicphoton}(right).


We want a reliable approximation to the non-perturbative cross section
so we can say where perturbative contributions dominate and where they
do not.  To get such an
approximation one can start with the cross section  given as

\begin{equation}
d\sigma(\gamma A \rightarrow \pi X) = \sum_V {\alpha \over \alpha_V}
            d\sigma(V + A \rightarrow \pi X) 
            + \Big({\rm non-VMD}\Big) ,
\end{equation}

\noindent where the sum is over vector mesons $V$, $\alpha = e^2/4\pi$,
and $\alpha_V = f_V^2/4\pi$.  We can get, for example, $f_\rho$ from
the decay  $\rho \rightarrow e^+e^-$.  Then ``all'' one needs is a
parameterization of the hadronic process, based on data.  Details of
our implementation of this program are given in~\cite{acw99}.

We took the soft processes to be polarization insensitive.  This agrees
with a recent Regge analysis of Manayenkov~\cite{m99}.


\begin{figure}


{\epsfxsize 2.7 in \epsfbox{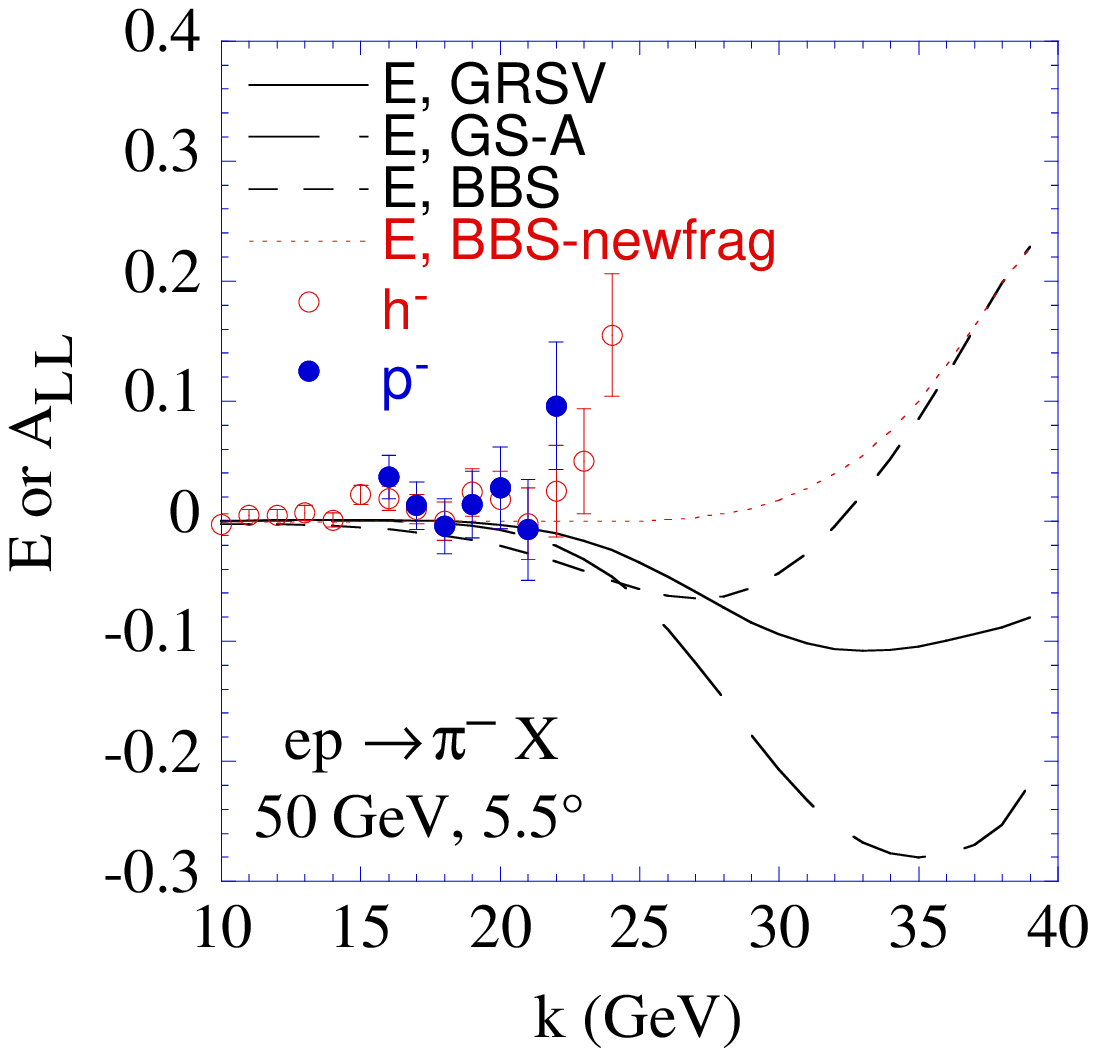}    }

{\vglue -2.44 in    \hglue 3in
\epsfxsize 2.55 in \epsfbox{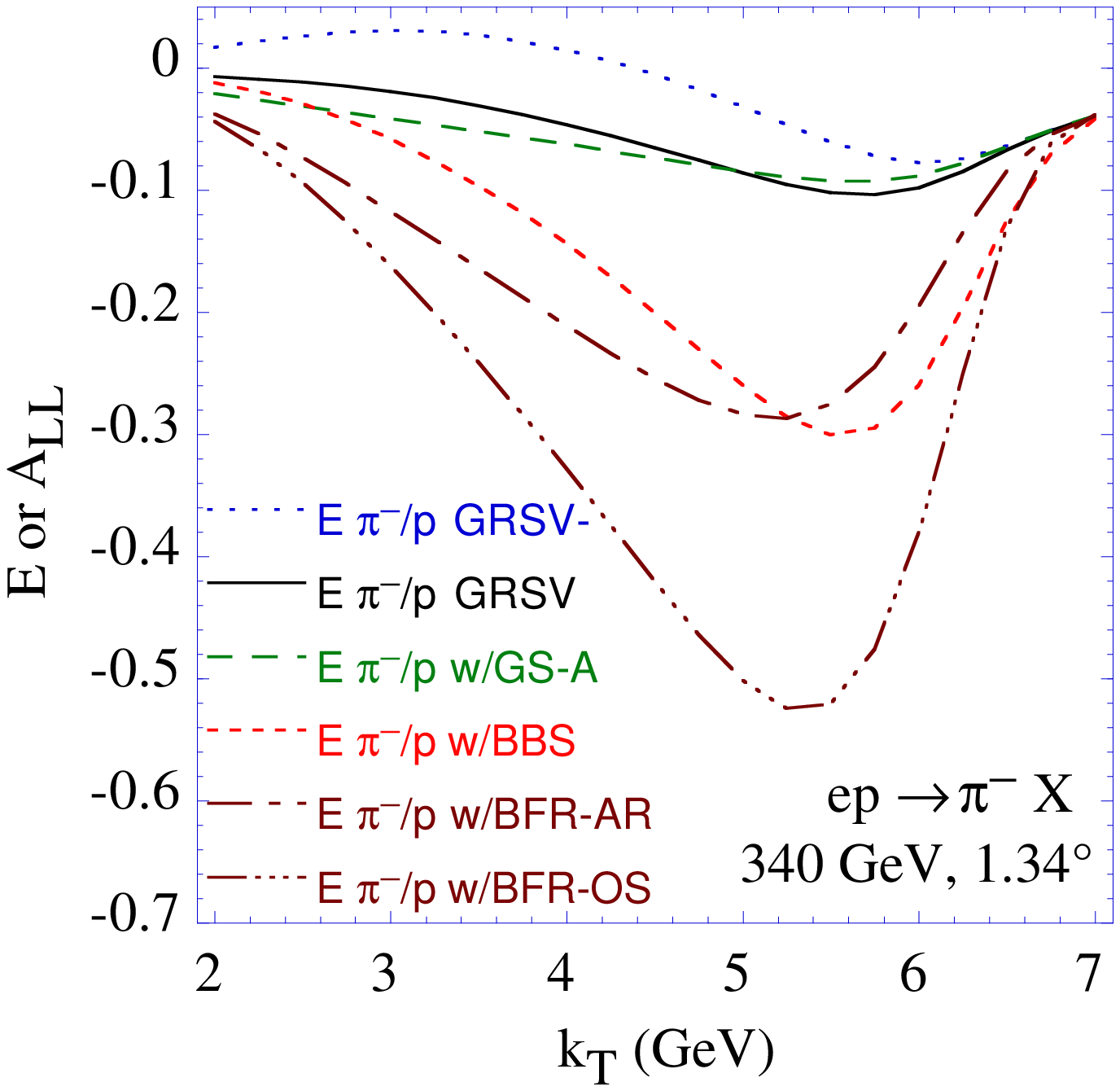}  }

\caption{Longitudinal polarization asymmetries for $\pi^-$
production for two energies and target
rest frame angles as labeled. A description of the curves is given in
the text. }   
\label{A_LL}

\end{figure}


\section*{Results}

Since results for the unpolarized cross section have already been
displayed in Fig.~\ref{sigma}, we focus on results for $A_{LL}$,
which is also called $E$ by some authors~\cite{barker75}. 
Fig.~\ref{A_LL} shows two plots, both for $\pi-$ production.

The 50 GeV plot, Fig.~\ref{A_LL}(left), is dominated by direct
pion production above the soft region, and is sensitive mainly to the
differing polarized quark distributions of the different models. 
Three different parton distribution
models are shown~\cite{bbs95,grsv96,gs96}.  Although the fragmentation
process is not the crucial one here, we should mention that mostly we
used our own fragmentation functions~\cite{cw93}, and that the results
using a better known set~\cite{bkk95} are not very different.  Neither
set of fragmentation functions agrees well with the most recent HERMES
data~\cite{makins} for unfavored vs. favored fragmentation functions,
and the one curve labeled ``newfrag'' is calculated with fragmentation
functions that agree better with that data.

Below about 20 GeV total pion momentum where the  soft process
dominates,  the data is well described by supposing the soft
processes are polarization independent.  Above that, with asymmetry
due to perturbative processes,  the difference
among the results for the different sets of parton distributions is
quite large for the $\pi^-$.

The data is from Anthony {\it et al.}~\cite{anthony99}. 
Presently most  of the data is in the region where the
soft processes dominate.  The data is already interesting.  Further
data at even higher pion momenta would be even more interesting. 
Regarding the differences among the quark distributions, recall
that large momentum corresponds to $x \rightarrow 1$ for the struck
quark,  and pQCD predicts that the quarks are 100\% polarized in this
limit.  Only the parton distributions labeled BBS~\cite{bbs95} are in
tune with the pQCD prediction, and they for large momentum predict
even a different sign for $A_{LL}$ for the $\pi^-$.  Calculated results
plotted with the data for the
$\pi^+$ and for deuteron targets may be examined in~\cite{acw99}.

The other plot in Fig.~\ref{A_LL} is for 340 GeV electron beam energy,
an energy where there is a long region where the fragmentation process
dominates. We would like to know how sensitive the possible
measurements of
$A_{LL}$ are to the different models for $\Delta g$.  To find out,
Fig.~\ref{A_LL} (right) presents calculated results for
$A_{LL}$ for one set of quark distributions and 5 different
distributions for
$\Delta g$~\cite{bbs95,grsv96,gs96,bfr96}.  The quark distributions and
unpolarized gluon distribution in each case are those of
GRSV~\cite{grsv96}.  There are 6 curves on each figure.  One of them
(labeled GRSV--) is a benchmark, which was calculated with 
$\Delta g$ set to zero.  The other curves use the $\Delta g$ from the
indicated distribution.  There is a fair spread in the results,
especially  for the $\pi^-$ where photon-gluon fusion gives a larger
fraction of the cross section.  Thus, one could adjudicate among the
polarized gluon distribution models.


\section*{Summary}                            \label{summary}


Several perturbative processes  contribute to hard pion
photoproduction.  All are calculable.  They give us new ways to measure
aspects of the pion wave function, and quark and gluon distributions,
especially $\Delta q$ and $\Delta g$.  The soft processes can be
estimated and avoided if the transverse momentum is greater than about
2 GeV.  SLAC or HERMES energies would be excellent for finding
direct pion production, which is sensitive to $\Delta u$ and $\Delta
d$, and higher energies would give a region where the fragmentation
process dominates and be excellent for measuring
$\Delta g$.


\section*{Acknowledgments}
My work on this subject has been done with Andrei Afanasev, Chris
Wahlquist, and A. B. Wakely and I thank them for pleasant
collaborations.  I have also benefited from talking to and reading the
work of many authors and apologize to those I have not explicitly 
cited.  I thank the NSF for support under grant PHY-9900657.


\end{document}